\documentclass[pre,preprint,amsmath,amssymb,floatfix]{revtex4}

\usepackage{times}
\usepackage{amsmath}
\usepackage{graphicx}
\usepackage{dcolumn}
\usepackage{bm}
\usepackage{psfrag}
\usepackage{slashbox}

\begin{document}

\preprint{APS/123-QED}

\title{Collision statistics in sheared inelastic hard spheres}

\author{Marcus~N.~Bannerman}
\author{Thomas E.~Green}
\author{Paul~Grassia}
\author{Leo~Lue}
\email{leo.lue@manchester.ac.uk}
\affiliation{
School of Chemical Engineering and Analytical Science \\
The University of Manchester \\
PO Box 88 \\
Sackville Street \\
Manchester \\
M60 1QD \\
United Kingdom
}

\date{\today}

\begin{abstract}
The dynamics of sheared inelastic-hard-sphere systems are studied
using non-equilibrium molecular dynamics simulations and direct
simulation Monte Carlo.  In the molecular dynamics simulations
Lees-Edwards boundary conditions are used to impose the shear.  The
dimensions of the simulation box are chosen to ensure that the
systems are homogeneous and that the shear is applied uniformly.
Various system properties are monitored, including the one-particle
velocity distribution, granular temperature, stress tensor,
collision rates, and time between collisions.
The one-particle velocity distribution is found to agree reasonably
well with an anisotropic Gaussian distribution, with only a slight
overpopulation of the high velocity tails.
The velocity distribution is strongly anisotropic, especially at
lower densities and lower values of the coefficient of restitution,
with the largest variance in the direction of shear.
The density dependence of the compressibility factor of the sheared
inelastic hard sphere system is quite similar to that of elastic
hard sphere fluids.
As the systems become more inelastic, the glancing collisions begin
to dominate more direct, head-on collisions.  Examination of the
distribution of the time between collisions indicates that the
collisions experienced by the particles are strongly correlated in
the highly inelastic systems.
A comparison of the simulation data is made with DSMC simulation of
the Enskog equation. Results of the kinetic model of Montanero et
al.\ {[}Montanero et al., J.\ Fluid Mech.\ 389, 391 (1999){]} based
on the Enskog equation are also included.  In general, good
agreement is found for high density, weakly inelastic systems.
\end{abstract}

\pacs{Valid PACS appear here}


\maketitle

\section{Introduction}

In rapid granular flows \cite{Campbell_1990,Goldhirsch_2003}, the mean
flight time of the particles in the granular material may be large
compared to the contact time between particles. Inter-particle
interactions are modelled as ``collisions,'' which play a key role in
transferring momentum and other properties through the system.
Granular materials in this flow regime can then be represented by a
collection of inelastic hard spheres
\cite{Campbell_Brennen_1985,Lutsko_2004b}.

The simplicity of the inelastic hard sphere model lends itself well to
theoretical analysis.  In particular, the methods developed for the
kinetic theory of equilibrium gases have been applied to rapidly
sheared inelastic hard sphere systems.  The seminal paper by Lun et
al.\ \cite{Lun_etal_1984} marked the start of ``complete'' kinetic
theories capable of predicting both the kinetic and collisional
properties.  The Boltzmann equation has featured predominantly in the
theory of granular gases due to its simpler form (e.g., see
Ref.~\cite{Santos_Astillero_2005}); however, the most successful
molecular kinetic theory to date is the revised Enskog
theory~\cite{ERNST_ETAL_1973_1}, an extension to the Boltzmann
equation for dense systems.  Enskog theory assumes uncorrelated
particle velocities and currently relies on a static structrual
correlation factor from elastic fluids
\cite{Carnahan_Starling_1969}.  Approximate theories beyond the Enskog
theory, such as ring theory \cite{Noije_etal_1998}, have been
developed and applied to granular systems, however, due to their
complexity, their use has been limited (e.g., cooling, rare granular
gases).

Common to most kinetic theory solutions is the assumption of a steady
state, spatially uniform distribution function.  Provided scale
separation exists, as is the case for elastic fluids, fluctuations
from this steady state can be accounted for using the Chapman-Enskog
expansion \cite{Sela_Goldhirsch_1998}.
To solve the Enskog equation, approximations typically begin by
taking moments of the kinetic equation with respect to the density,
velocity, and products of the velocity.  These moment equations are
used to solve for the parameters of an expansion or model.
Typically, only terms up to the granular ``temperature'', or isotropic
stress and rotation terms \cite{Lun_1991}, are included as field
variables.  Anisotropic stresses can still be predicted from such a
theory \cite{Goldhirsch_2008}; indeed, attempts have been made to
include the full second order velocity moment \cite{JENKINS_1988} as a
hydrodynamic variable to improve theoretical predictions.

Grad's method \cite{Jenkins_Richman_1985} solves Enskog theory using
an expansion of the distribution function about a reference
state.  This has been applied to poly-disperse granular systems
\cite{Lutsko_2004b} and, unlike perturbative solutions, does not require
assumptions on the strength of the shear.
Kinetic models are a powerful method of generating simplified kinetic
equations which retain key features of the original.  Montanero et
al.\ \cite{Montanero_etal_1999} solved an improved
Bhatnagar-Gross-Krook (BGK) kinetic model
\cite{Santos_etal_1998,Brey_etal_1999} for inelastic systems.  The
improved BGK model approximates the collisional term of the kinetic
equation using the first two velocity moments, which correspond to the
collisional stress and energy loss, and a general relaxation term.
This leads to a simplified kinetic equation.  The solution in the low
dissipation limit is particularly attractive, as it provides estimates
for the system properties without requiring numerical solution and
compares favourably to Direct Simulation Monte Carlo (DSMC) results.

DSMC \cite{bird_1994} is a numerical simulation technique to directly
solve the Boltzmann equation without requiring further approximations.
This can then be used to rapidly test solutions of the kinetic
equation.  The method has already been extended to the Enskog equation
for homogeneously sheared inelastic systems
\cite{Hopkins_Shen_1992,Montanero1997,Montanero_etal_1999}.

While kinetic theories do offer insight into the behavior of granular
materials, they are necessarily approximate. 
The Boltzmann and Enskog kinetic theories do not include velocity or
dynamic structural correlations.  Ring theory \cite{Noije_etal_1998}
is capable of including particle correlations; however, further
approximations are required to make the resulting theory tractable.
These correlations are present in moderately dense to dense systems of
elastic particles, but they are enhanced by the clustering in
inelastic systems \cite{Alam_Luding_2003,Alam_Luding_2005}.  The
failure of Boltzmann and Enskog theories at high densities is
therefore expected, even for elastic hard sphere systems.
On the other hand, non-equilibrium molecular dynamics (NEMD)
simulations \cite{Campbell_1989,Campbell_Brennen_1985} can, in
principle, give ``exact'' results for driven inelastic-hard-sphere
systems \cite{Herbst_etal_2004,Herrmann_etal_2001}.  These simulations
can be used to validate kinetic theories against the underlying model.
Initial studies of sheared granular systems used moving boundaries
\cite{Campbell_Brennen_1985,Conway_Glasser_2004}, such as rough walls,
to introduce energy into the system.  Due to the computational
limitations, the wall separation is typically of the order of a few
particle diameters, and wall effects dominate the simulation results.
For large system sizes, shear instability is observed
\cite{Alam_etal_2005}.  Consequently, the results for wall driven
simulations are strongly dependent on system size.

Another manner to introduce shear in non-equilibrium molecular
dynamics is the Lees-Edwards \cite{Lees_Edwards_1972} or ``sliding
brick'' boundary conditions.  Simulations of inelastic hard-sphere
systems using Lees-Edwards boundary conditions
\cite{Walton_Braun_1986a,Campbell_1989,Hopkins_Louge_1991,Goldhirsch_Tan_1996,Tan_Goldhirsch_1997}
lessen the influence of wall effects, by elimination of the surface of
the system, but these simulations still introduce shear in an
inhomogenous manner, which leads to clustering instabilities
\cite{Lutsko_2004} for larger systems.

While there are many interesting similarities between elastic
hard-sphere fluids and driven inelastic hard-sphere systems, there are
key differences.  One is the tendency of inelastic hard spheres to
form clusters and patterns, while elastic hard sphere fluids tend
to remain isotropic.  Another example is the velocity distribution.
The velocity of elastic hard spheres is governed by the Maxwell
distribution, which is isotropic and Gaussian.  The velocity
distribution of flowing inelastic hard spheres is, in general
anisotropic \cite{Losert_etal_1999}, and can show significant
deviations from the Gaussian distribution, especially when there is
clustering.

In this work, we examine the properties of sheared
inelastic-hard-sphere systems using non-equilibrium event-driven
molecular dynamics simulations with the SLLOD algorithm combined with
Lees-Edwards boundary conditions.  Part of the purpose of this work is
to investigate, at a particle level, the differences between the
behavior of inelastic and elastic (equilibrium) hard sphere systems.
Another purpose of this work is to provide simulation data which can
be used to test kinetic theory predictions for the properties of these
systems. 
A previous study by Montanero et al.\ \cite{Montanero_etal_2006} has
already compared 2D and 3D simulations of binary inelastic hard
spheres against DSMC simulations of Enskog theory.  They find good
agreement over the range of mass ratio, size ratio and inelasticity
studied; however, the clustering instability present in systems of
large numbers of highly inelastic particles appears to limit the range
of inelasticity studied.
As mentioned previously, kinetic theories for sheared granular
materials are typically developed for the case where the system is
spatially uniform and homogeneously sheared. 
One of the difficulties with comparing the predictions of the kinetic
theory with the simulation data for sheared granular materials is the
formation of clusters, which makes comparison between the two
problematic.  As a consequence, care is taken in this work to ensure
that the systems remain homogeneous and strongly inelastic systems can
be accessed.
In these simulations, we investigate the collision statistics, such as
velocity distributions, collision angles, time between collisions and
mean free paths, of sheared inelastic hard spheres.  In addition, we
examine the variation of various bulk properties of the system, such
as the viscosity, mean kinetic energy, and stress, with the packing
fraction and coefficient of restitution of the particles.
We also investigate the correlations between the collisions, which are
neglected in most kinetic theory approaches.
The remainder of this paper is organized as follows.  In
Section~\ref{sec:simulation}, we describe the details of the granular
dynamics simulations. In Section~\ref{sec:DSMC}, we describe the
details of the DSMC simulations. In Section~\ref{sec:results}, we
present the results of our simulation work, including a comparison
with the predictions of Enskog theory.  Finally, a summary of the main
findings is provided in Section~\ref{sec:conclusions}.

\section{
Simulation details
\label{sec:simulation}}

Non-equilibrium granular dynamics simulations were performed on
systems of inelastic hard spheres of diameter $\sigma$ and mass $m$.
The system is sheared in the $y$-plane in the $x$-direction with a
constant strain rate of $\dot{\gamma}$ using the SLLOD
algorithm~\cite{EVANS_MORRIS_1990}.  In this method, shear is applied
through the use of the Lees-Edwards sliding brick boundary
conditions~\cite{Lees_Edwards_1972,EVANS_MORRIS_1990} and the velocity
is transformed relative to a linear velocity profile.  The equations
of motion are
\begin{align}
\label{eq:SLLODr}
\frac{d{\bf r}_{i}}{dt} =& \bar{\bf v}_{i} 
+ y_{i} \dot{\gamma}\hat{\bf e}_x
\\
\frac{d\bar{\bf v}_{i}}{dt} =& \frac{{\bf F}_{i}}{m}
-\bar{v}_{y,i} \dot{\gamma}\hat{\bf e}_x
\label{eq:SLLODv}
\end{align}
where ${\bf F}_i$ is the force acting on particle $i$, ${\bf r}_{i}$
is the position of particle $i$, $\bar{\bf v}_{i}$ is the so called
peculiar velocity of particle $i$, $y_{i}$ is the $y$-coordinate of
particle $i$, $\bar{v}_{y,i}$ is the $y$-component of the peculiar
velocity of particle $i$, $\hat{\bf e}_x$ is a unit vector pointing in
the positive $x$-direction, and $\dot{\gamma}$ is the strain rate.

The peculiar velocity of a particle $i$ is defined as the difference
between its lab velocity ${\bf v}_i$ and the local streaming velocity
(the velocity of the local streamline).  For simple shear, it is
given by the linear transformation
\begin{align*}
\bar{\bf v}_i = {\bf v}_i - y_{i}\dot{\gamma}_i\hat{\bf e}_x
\end{align*}
The peculiar velocity is related to the dispersion of the particles
from the average streamlines of the flow.  The SLLOD equations of
motion are particularly convenient as the peculiar velocity is
naturally recovered without the need for a separate co-ordinate
transformation.  They allow the possibility of thermostatting the
system~\cite{Petravic_Jepps_2003} and the study of time
dependent shear flows.

In a hard-sphere system, the spheres do not experience any force
between collisions.  The equations of motion can then be solved
analytically for the trajectories of the spheres between collisions.
The evolution of the position and peculiar velocity of particle $i$ in
the system between collisions is
\begin{align}
{\bf r}_{i}(t) =& {\bf r}_{i}(t_0) 
+ [\bar{\bf v}_i(t_0) + y_i(t_0)\dot{\gamma}\hat{\bf e}_x](t-t_0)
\nonumber
\\
=& {\bf r}_{i}(t_0) + {\bf v}_i(t_0)(t-t_0)
\nonumber
\\
\bar{\bf v}_{i}(t) =& \bar{\bf v}_{i}(t_0) 
- \bar{v}_{y,i}(t_0) \dot{\gamma}(t-t_0) \hat{\bf e}_x
\label{eq:trajectory}
\end{align}

When a particle undergoes a collision, it experiences an impulse which
alters its velocity.  These collisions are instantaneous and only
occur between pairs of spheres (i.e., there are no three or higher
body collisions).  The inelasticity of the hard spheres is
characterized by the coefficient of restitution $\alpha$. This is
defined through the amount of kinetic energy $\Delta{E}$ lost on
collision
\begin{equation}
\Delta{E} = \frac{m}{4}(1-\alpha^2)
\left(\hat{\bf r}_{ij}\cdot{\bf v}_{ij}\right)^2
\end{equation}
where ${\bf v}_i$ is the velocity of particle $i$ immediately before
collision, ${\bf v}_{ij}={\bf v}_i-{\bf v}_j$, and $\hat{\bf
  r}_{ij}={\bf r}_{ij}/|{\bf r}_{ij}|$ is the unit vector pointing
from the center of particle $j$ to the center of particle $i$.

Each collision preserves the total momentum of the particles involved;
therefore, the change of velocities for a colliding pair of spheres
$i$ and $j$ is given by
\begin{align}
{\bf v}_{i}' =& {\bf v}_{i} 
- \frac{1}{2}(1+\alpha)
  \left(\hat{\bf r}_{ij} \cdot {\bf v}_{ij}\right)
  \hat{\bf r}_{ij}
\nonumber
\\
{\bf v}_{j}' =& {\bf v}_{j} 
+ \frac{1}{2}(1+\alpha)\left(\hat{\bf r}_{ij}\cdot{\bf v}_{ij}\right)
  \hat{\bf r}_{ij}
\label{eq:change_in_vel_inelastic}
\end{align}
where the primes denote post collision values of the particle
velocities.

The coefficient of restitution $\alpha$ is, in general, a function of
the relative velocity on collision.  Viscoelastic models that
incorporate this have been very successful in describing real systems
such as steel spheres \cite{McNamara_Falcon_2003}. A common
approximation in kinetic theory is to assume a constant coefficient of
inelasticity, as this greatly simplifies the collision integrals while
the basic physics is not significantly altered.  A constant
coefficient of restitution is used in this work to facilitate
comparison against kinetic theory results.

One concern for a system with a constant coefficient of restitution is
the phenomenon of inelastic collapse, where an infinite number of
collisions occur between several spheres in a finite interval of time.
Event-driven simulations will fail in the event of a single collapse
event.  In two dimensional, freely cooling, inelastic hard sphere
systems undergo \cite{McNamara_Young_1996} inelastic collapse with
coefficient of restitution as high as 0.59.

Inelastic collapse is rare in sheared systems \cite{Alam_Hrenya_2001}
and is increasingly rare in higher dimensions; however, a near
collapse situation can still cause a simulation to break down if the
machine precision is not sufficiently high to resolve a rapid series
of successive collisions.  
In the simulations performed in this work, no partial or full collapse
events were found, even for dense and highly inelastic systems.

The simulation algorithm that we employ is a generalization of the
standard event-driven molecular dynamics algorithm for hard
spheres~\cite{Alder_Wainwright_1959} (see
Ref.~\cite{Bannerman_Lue_Site}).  The main modifications are the use
of the sliding brick boundary conditions \cite{Lees_Edwards_1972} and
the SLLOD equations of motion.

Unlike the elastic hard sphere system, the inelastic hard-sphere
system has no intrinsic time scale.  The applied strain rate
$\dot{\gamma}$ sets the time scale of the system.  Therefore, there
are only two relevant dimensionless parameters: the density
$\rho\sigma^3$ and the coefficient of restitution $\alpha$.
In this work, the density is varied from $\rho\sigma^3=0.4$ to $0.9$,
and the coefficient of restitution is varied from $\alpha=0.4$ to
$0.9$.

\begin{figure}
\centering
\includegraphics[clip,width=\columnwidth]{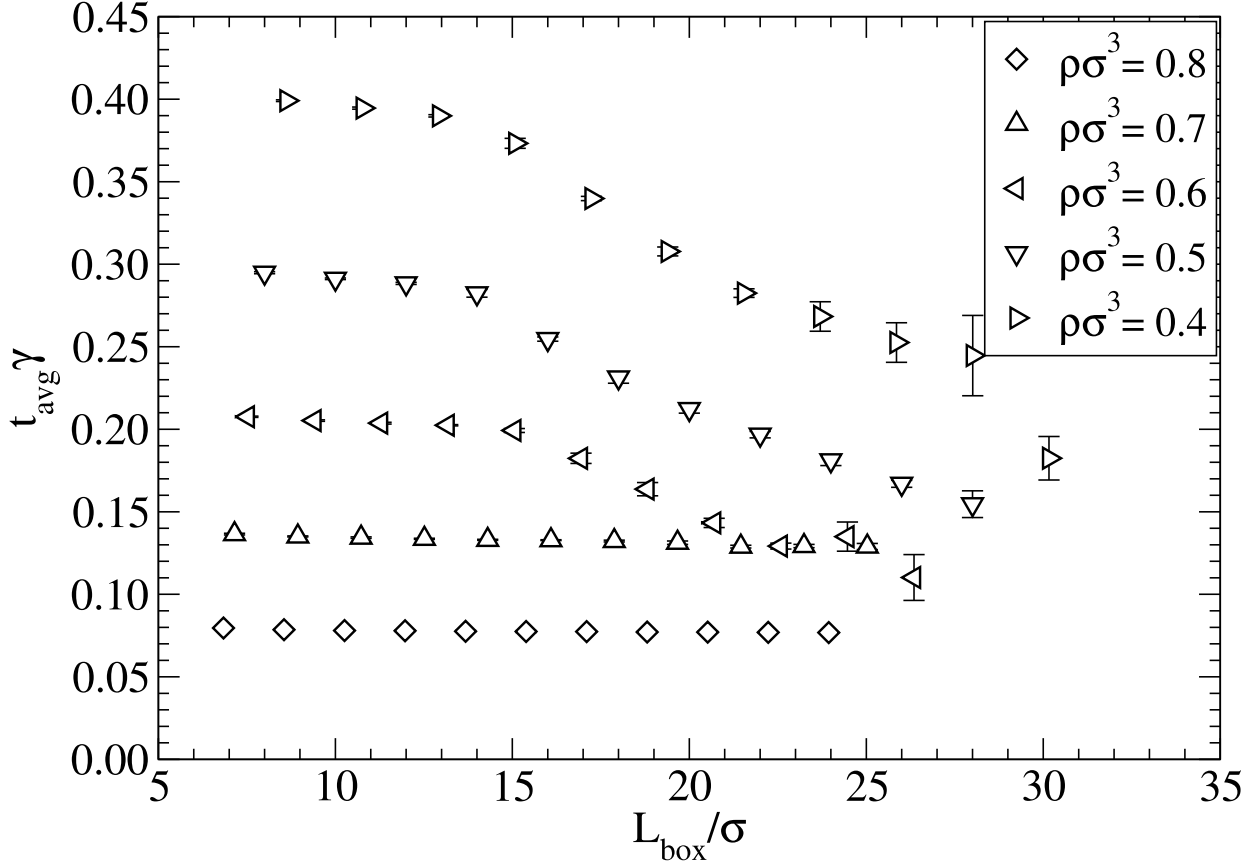}
\caption{\label{fig:MFTNScaling} The system size dependence of the
  mean free time between collisions ${\rm t}_{\rm avg}$ for
  simulations of a sheared, inelastic hard sphere system with
  $\alpha=0.4$ in a cubic box with sides of length $L_{\rm box}$. The
  number of particles is $N=256$ in the smallest system and
  $N=10,976$ in the largest.}
\end{figure}

Because the shear is imposed through the boundary conditions, the
strain rate is only fixed at two points, separated by the entire
height of the simulation box.  In low density systems with large
numbers of particles, clustering occurs \cite{Alam_etal_2005}.  This
leads to a local variation of the strain rate in the system, and,
consequently, the system will not be homogeneously sheared.
At the onset of clustering, the size dependence of the system
properties changes from the typical $N^{-1}$ scaling to a different
behavior.  To illustrate this, the mean free time of sheared
inelastic-hard-spheres with $\alpha=0.4$ is shown in
Fig.~\ref{fig:MFTNScaling}.  These simulations were performed in a
cubic simulation box where the system size was varied while holding
the density constant.  The ``break'' in the curves for the lower
densities indicates the presence of cluster formation in the larger
systems.  The same general behavior manifests in all system properties
and is relatively easy to detect.

Kinetic theory studies of sheared inelastic systems typically assume
that the system is homogeneous and uniformly sheared, with a linear
velocity profile.  This makes the comparison between granular dynamics
simulations and the kinetic theory problematic.  To allow comparison
with these theories, we ensure that the systems remain homogeneous
during the course of the simulations.
In order to avoid the clustering regime while still maintaining a
large system size to provide proper statistics, the $x$-, $y$-, and
$z$-dimensions of the simulation box are set to the ratio $14.4:1:1$
and there are a total $N=7200$ spheres.  This ensured that the
systems remained homogeneous for all conditions (i.e.\ number of
particles, coefficient of restitution, and density) that were examined
in this work.

At the beginning of the simulations for each set of conditions, the
spheres are arranged in a face centered cubic lattice at the
appropriate density.  The velocities of the spheres are initially
assigned from a Maxwell-Boltzmann distribution.  The simulations are
then run for an ``equilibration'' period of $10^7$ collisions.
Afterwards, system property data are collected over at least $10$
production runs, each lasting $10^7$ collisions.  The uncertainties of
the data are estimated from the standard deviations of the results
from these separate runs.  In the next section, we describe the DSMC
simulations performed.

\section{\label{sec:DSMC}DSMC Simulations}

The DSMC method was used to numerically solve the Enskog equation.
This technique has been described in detail previously
\cite{Hopkins_Shen_1992,Montanero_etal_1999} and is only covered
briefly here.  The peculiar velocity distribution function $f$ is
represented by using a collection of $N$ sample velocities or
``simulated'' particles:
\begin{align}
f(\bar{\bf v}, t)= {N}^{-1} 
\sum_{i=1}^{{N}} 
  \delta^3\left(\bar{\bf v}-\bar{\bf v}_i(t)\right),
\end{align}
where $\bar{\bf v}_i(t)$ is the peculiar velocity of sample $i$ at
time $t$.  At each time step $\Delta t$, the samples are evolved
according to the SLLOD dynamics (see Eq.~(\ref{eq:SLLODv})).  The
samples are then tested for collisional updates.  
At each time step, $\frac{1}{2}{N}P_{\rm max}^{(c)}$ pairs of
samples in the collection are selected, where $P_{\rm max}^{(c)}$ is a
parameter of the DSMC simulation.  The probability that a collision
between a pair of samples $i$ and $j$ will be executed is proportional
to
\begin{align}
P_{ij}^{(c)}=4\pi\sigma^2\chi \rho \left(\hat{\bf k}
  \cdot {\bf v}_{ij}\right) 
  \Theta\left(\hat{\bf k}\cdot {\bf v}_{ij}\right)\Delta t
\end{align}
where $\hat{\bf k}$ is a randomly generated unit vector,
${\bf{}v}_{ij}=\bar{\bf{}v}_i-\bar{\bf{}v}_i-\sigma\dot\gamma\hat{k}_y\hat{\bf{}e}_x$
is the relative lab velocity, $\Theta$ is the Heaviside step function,
and $\chi$ is the radial distribution function at contact.  In this
work, the value of $\chi$ is taken from the Carnahan-Starling
\cite{Carnahan_Starling_1969} equation of state for elastic hard
spheres, which is given by
\begin{align}
  \chi = \frac{1-\nu/2}{\left(1-\nu\right)^3}
\end{align}
where $\nu=\rho\pi\sigma^3/6$ is the solid fraction.

To optimize the simulation, the quantity $P_{\rm max}^{(c)}$ is chosen
to be the maximum observed value of $P_{ij}^{(c)}$. This is estimated
and updated during a simulation if $P_{ij}^{(c)}$ exceeds
$P_{max}^{(c)}$. The probability that a collision between samples $i$
and $j$ is executed is $P_{ij}^{(c)}/P_{\rm max}^{(c)}$, and, if the
collision is accepted, the velocities are updated using
Eq.~(\ref{eq:change_in_vel_inelastic}) with ${\bf r}_{ij} =
-\sigma\hat{\bf k}$.

For the results presented here, ${N}=1372$ and $\Delta t$ is
selected such that $\frac{1}{2}{N}P_{\rm max}^{(c)} < 5$.
The distribution functions are equilibrated for $10^6$ collisions, and
then results are collected and averaged over $10$ runs of $10^7$
collisions.

\section{Results and discussion
\label{sec:results}}

In this section, we present our simulation results for the properties
of homogeneously sheared inelastic hard sphere systems.  These results
are compared against DSMC simulation of the Enskog equation to test
the Enskog approximation.  We also include the results from the
kinetic model solved by Montanero et al.\
\cite{Dufty_etal_1997,Montanero_etal_1999}. This theory is
particularly interesting as it provides analytical results in the
limit of small strain rates, along with simple expressions that
approximate DSMC results. Without the small strain rate approximation,
a more accurate numerical solution of the model is available
\cite{Montanero_etal_1999}; however, the DSMC simulations already
provide accurate Enskog theory results without further approximation.
%

\subsection{Velocity distribution}

The kinetic energy of the system is defined through the fluctuations
of the velocity of the particles from their respective local streaming
velocity
\begin{equation}
{E} = \frac{1}{2}\sum_{k=1}^N m \bar{v}_k^2.
\end{equation}
The mean kinetic energy is therefore a measure of the velocity
dispersion present in the system.  In analogy with elastic
(equilibrium) hard sphere fluids, a kinetic (or ``granular'')
temperature $T$ is typically introduced through the relation
\begin{equation}
\frac{3}{2} N k_BT \equiv \langle{E}\rangle,
\end{equation}
where $N$ is the number of particles in the system, and $k_B$ is the
Boltzmann constant.  Although the physical significance of the
``granular temperature'' has been a subject of some controversy
\cite{Goldhirsch_2008}, the concept has proved effective in the
theoretical modelling of the properties of granular materials.

\begin{figure}
\centering
\includegraphics[clip,width=\columnwidth]{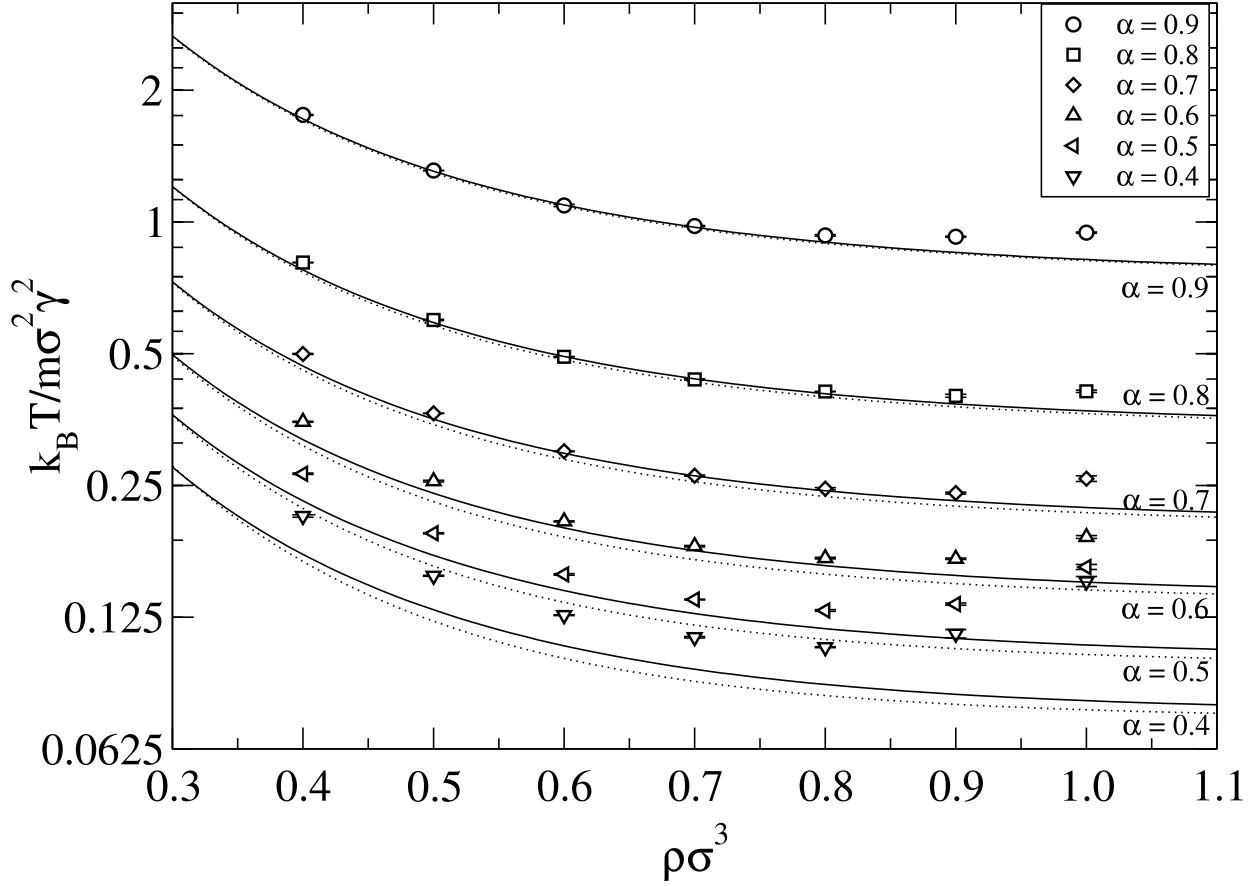}
\caption{\label{fig:ke}
Variation of the mean kinetic energy per particle with density for
(i) $\alpha=0.4$ (circles),
(ii) $\alpha=0.5$ (squares),
(iii) $\alpha=0.6$ (diamonds),
(iv) $\alpha=0.7$ (triangles-up),
(v) $\alpha=0.8$ (triangles-left), and
(vi) $\alpha=0.9$ (triangles-down).
The dotted lines are the suggested expressions of
Ref.~\cite{Montanero_etal_1999} and the solid lines are DSMC results.}
\end{figure}

The granular temperature of the sheared inelastic hard sphere system
at steady state is plotted in Fig.~\ref{fig:ke}.  The symbols are the
results of our molecular dynamics simulations, the dotted lines are
the suggested expressions of Montanero et al.\
\cite{Montanero_etal_1999}, and the solid lines are the DSMC
simulation results.
From the figure, it can be seen that the granular temperature of the
system decreases with decreasing values of the coefficient of
restitution.  The particles in a strongly inelastic system rebound
less from collisions; therefore, collisions in the direction of shear
can quickly settle a particle to the velocity of the streamline.  In
addition, the motion of the particles off the streamline (in the $y$-
and $z$-directions) are more quickly dissipated by collisions with
particles on neighboring streamlines.  Consequently, strongly
inelastic systems have a greater tendency to follow the streamlines of
a flow.

At low densities, the granular temperature increases with decreasing
particle densities.  The collisions between particles transmit
information regarding the mean velocity of the flow.  For very low
density systems, the collisions are relatively rare events, and
between collisions a particle will generally travel on trajectories
that deviate from the streamlines, thus contributing to the granular
temperature.  With increasing density, a particle will become
increasingly ``caged'' by surrounding particles, experiencing more
collisions that will keep it on a particular streamline.  Therefore,
one expects that the temperature should generally decrease with
increasing particle density.
However, the simulation data indicate that the temperature of the
system does not depend monotonically with the density, and a minimum
is observed at a relatively high density for all the systems
considered.  The minimum becomes more pronounced as the coefficient of
restitution decreases.

We note that in dense experimental granular systems, particles mainly
remain in contact with each other and interact by rolling or sliding
past one another, rather than through collisions.  In this regime,
soft sphere models \cite{Campbell_2002}, as opposed to hard sphere
models, are more representative.
Consequently, the applicability of the simulation results for the
inelastic hard sphere system at high densities to experimental
granular systems should be considered with care.

In general, Enskog theory and the solution of Montanero et al.\
provides a fairly accurate description of the simulation results;
however, there is a large discrepancy for high values of the
inelasticity and density.  In addition, Enskog theory does not capture
the presence of the minimum in the temperature with respect to the
density.

\begin{figure}
\begin{center}
\includegraphics[clip,width=\columnwidth]{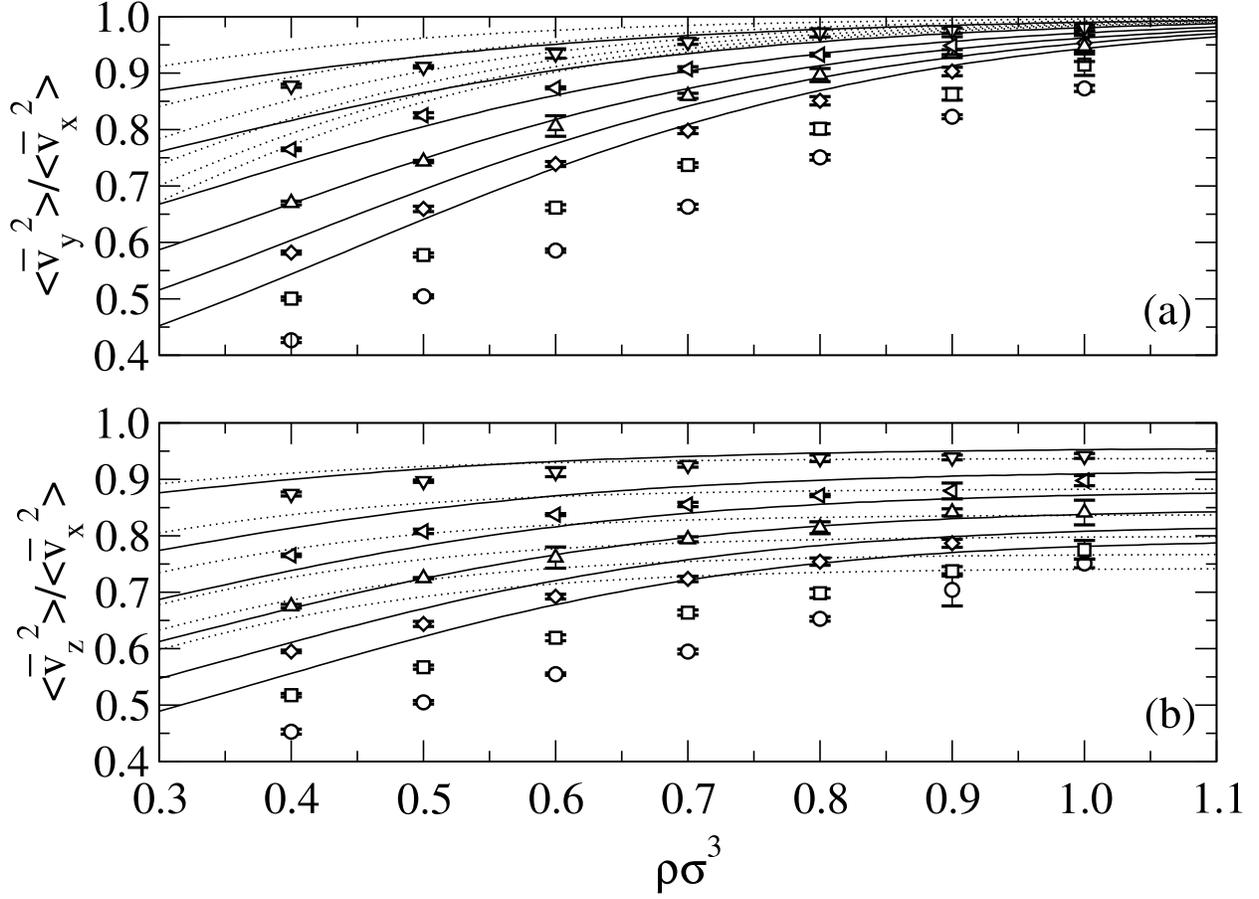}
\caption{\label{fig:ratio}
The ratio of the mean-squared-velocity (a) in the $y$- and
$x$-directions and (b) in the $z$- and $x$-directions for sheared
inelastic hard sphere sytems with
(i)   $\alpha=0.4$ (circles),
(ii)  $\alpha=0.5$ (squares),
(iii) $\alpha=0.6$ (diamonds),
(iv)  $\alpha=0.7$ (triangles-up),
(v)   $\alpha=0.8$ (triangles-left), and
(vi)  $\alpha=0.9$ (triangles-down).
The dotted lines are the suggested expressions of
Ref.~\cite{Montanero_etal_1999}, and the solid lines are the DSMC results.}
\end{center}
\end{figure}

Equilibrium fluids obey the equipartition theorem: energy is, on
average, distributed evenly between all degrees of freedom.  In driven
granular systems, however, this has been shown to not be the case
\cite{Knight_Woodcock_1996}.
%
%
Figure~\ref{fig:ratio}a shows the variation of $\langle
\bar{v}_y^2\rangle/\langle \bar{v}_x^2\rangle$ with density, and
Fig.~\ref{fig:ratio}b shows the variation of
$\langle{\bar{v}_z^2\rangle/\langle \bar{v}_y^2}\rangle$.  The symbols
are the results of our molecular dynamics simulations, the dotted
lines are the predictions of the theory of Montanero et
al.\ \cite{Montanero_etal_1999}, and the solid lines are the DSMC
simulation results.
If the system obeyed the equipartition function, then both these
ratios would be equal to one.  The dispersion of the velocity parallel
to the direction of shear (i.e.\ the $x$-direction) is consistently
larger than that perpendicular to the shear, which is unsurprising as
this is the direction in which energy is inputted to the system.  The
asymmetry increases with decreasing density and with decreasing values
of the coefficient of restitution.  It is interesting to note,
however, that the fluctuations in the velocity in the $y$- and
$z$-directions are nearly equal.

The low dissipation theory of Montanero et al.\ strongly under
predicts the anisotropy in the velocity dispersion. DSMC results
provide a better description but still deviate significantly from the
simulation results at low values of the elasticity.

Montanero et al.'s theory truncates terms within the second velocity
moment of the collision integral and all higher terms. The full second
moment could be included to improve predictions; however, as this is
primarily a collision term it is unlikely to improve the predictions
of the velocity anisotropy.  

The kinetic model could be expanded by relaxing to a generalized
Gaussian distribution, as in the ellipsoidal statistical model. The
extra degrees of freedom in the model would then be solved for by the
inclusion of a full second velocity moment balance. This might still
prove tractable and improve the predictions for the velocity
dispersion anisotropy.

\begin{figure}
\begin{center}
\includegraphics[clip,width=\columnwidth]{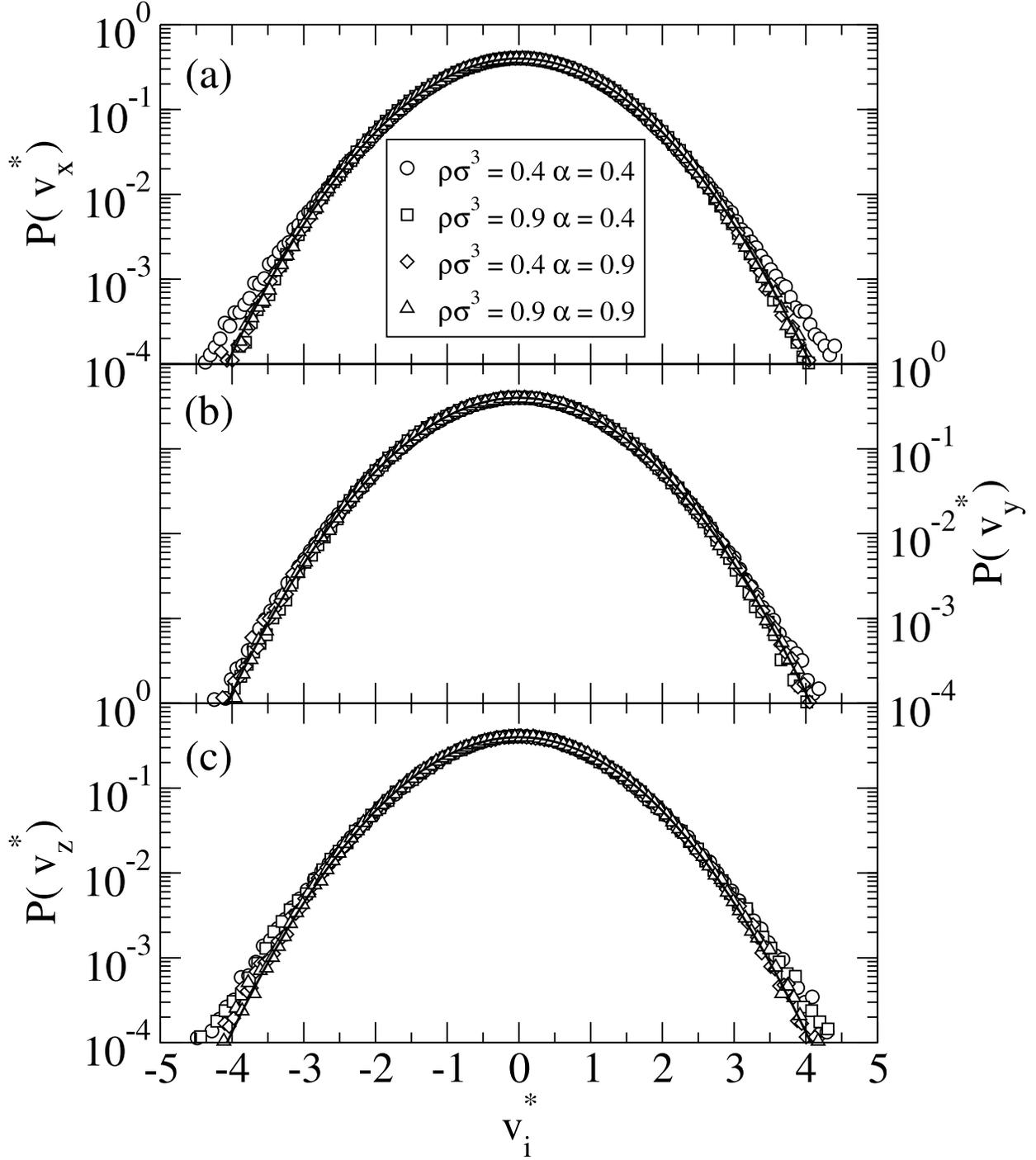}
\caption{\label{fig:vel} 
The peculiar velocity distribution in the 
(a) $x$-direction, 
(b) $y$-direction, and 
(c) $z$-direction 
in sheared inelastic hard sphere systems with
(i) $\rho\sigma^3=0.4$ and $\alpha=0.4$ (circles), 
(ii) $\rho\sigma^3=0.9$ and $\alpha=0.4$ (squares),
(iii) $\rho\sigma^3=0.4$ and $\alpha=0.9$ (diamonds), and
(iv) $\rho\sigma^3=0.9$ and $\alpha=0.9$ (triangles).
The solid line is a Gaussian distribution.}
\end{center}
\end{figure}

For equilibrium systems, such as elastic-hard-sphere fluids, the
velocity distribution is exactly given by the Maxwell-Boltzmann
distribution; however, granular materials have been shown to deviate
from this distribution
\cite{Olafsen_Urbach_1999,Losert_etal_1999,Rouyer_Menon_2000}.  
The simulation data for the distributions of the single particle $x$-,
$y$-, and $z$-component of the peculiar velocity are shown in
Fig.~\ref{fig:vel}.  The peculiar velocities are reduced by their mean
squared values ($v_i^*=\bar{v}_i/\left\langle
\bar{v}_i^2\right\rangle^{1/2}$, for $i=x$, $y$, and $z$).
The distributions are, in general, well described by an anisotropic
Gaussian distribution.  For the highly inelastic systems, the
distributions display a slightly enhanced high velocity tail.  This is
most evident in the direction of shear (i.e.\ the $x$-direction).

For simulations in a {\em cubic} box at the onset of clustering in the
system, the peculiar velocity distributions in the $y$- and
$z$-directions can be shown to develop strong high velocity tails.  In
this case, the bulk of the particles are within a dense low strain
rate zone, while the remainder reside in a rare, high strain rate and
granular temperature region.  The particles in the high strain rate
region lead to a high velocity tail.  Further studies on clustering
effects are currently underway.

\subsection{Stress tensor}

In this section, we examine the stress tensor.  The time averaged
value of the stress tensor $\langle {\bf P}\rangle$ for a hard-sphere
system is given by \cite{Alder_etal_1970}
\begin{equation}
\langle {\bf P}\rangle
= \frac{1}{V\tau}\sum_{{\rm collisions}}^{\tau}\left[
\Delta t_{c} \sum_{k=1}^{N}m\bar{\bf v}_{k}\bar{\bf v}_{k}
+ \sigma\hat{\bf r}_{ij}m\Delta{\bf v}_i
\right]
\label{eq:stress_tensor}
\end{equation}%
where $\Delta t_c$ is the time interval between two consecutive
collisions, $\Delta{\bf v}_i$ is the change of velocity of sphere $i$
on collision, $\tau$ is the time over which the stress tensor is
averaged, and $V$ is the total volume of the system.  The first
summation runs over all collisions that occur during the time $\tau$,
and the indexes $i$ and $j$ refer to the spheres undergoing the
collision; the index $k$ runs over all particles in the system.

\begin{figure}
\begin{center}
\includegraphics[clip,width=\columnwidth]{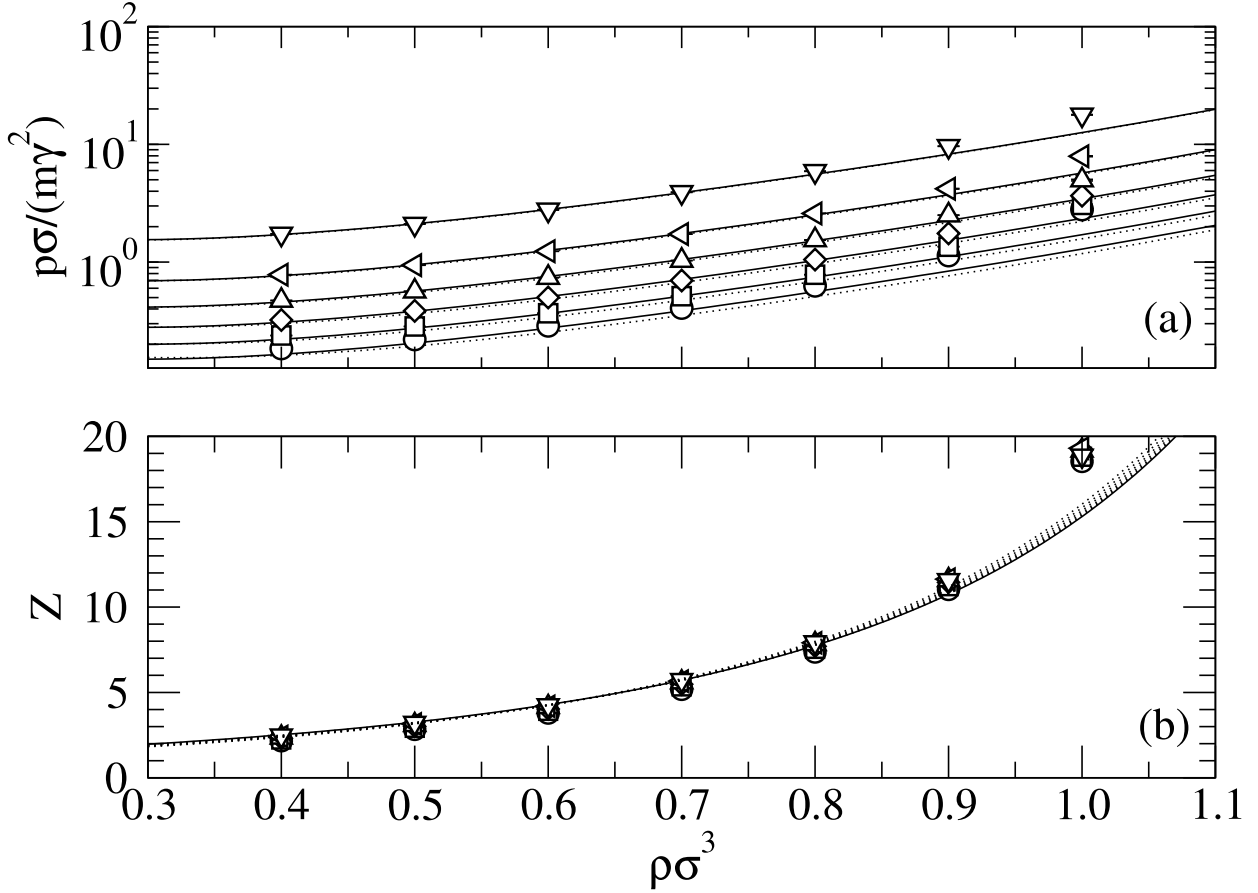}
\caption{\label{fig:pressure} 
The (a) dimensionless pressure $p\sigma/(m\dot{\gamma}^2)$ and (b) the
compressibility factor $Z$ for inelastic hard-sphere systems with
(i)   $\alpha=0.4$ (circles),
(ii)  $\alpha=0.5$ (squares),
(iii) $\alpha=0.6$ (diamonds),
(iv)  $\alpha=0.7$ (triangles-up),
(v)   $\alpha=0.8$ (triangles-left), and
(vi)  $\alpha=0.9$ (triangles-down).  
The uncertainty is smaller than the symbol size.  The dotted lines are
the suggested expressions of Ref.~\cite{Montanero_etal_1999}, and the
solid lines are DSMC results.  The solid line in (b) is the
Carnahan-Starling equation of state for elastic hard sphere fluids,
and the dotted lines are DSMC results for various $\alpha$.}
\end{center}
\end{figure}

The pressure $p$ of the system, which is defined as
\begin{equation*}
p \equiv \frac{1}{3} \left(
\langle P_{xx}\rangle+\langle P_{yy}\rangle+\langle P_{zz}\rangle
\right),
\end{equation*}%
is plotted in Fig.~\ref{fig:pressure}a.  As expected, the pressure
increases as the density of the system increases.  It also increases
with increasing coefficient of restitution due to the rise in granular
temperature.  
The Enskog theory requires, as input, the collision rate between
particles as a function of the density.  This is typically given by
the equation of state for elastic hard sphere fluids through the
compressibility factor.
The compressibility factor $Z$, defined as
\begin{equation*}
Z \equiv \frac{p}{\rho k_BT},
\end{equation*}%
is plotted for the shear inelastic-hard-sphere system in
Fig.~\ref{fig:pressure}b.  The symbols represent the results of the
simulations, and the line is the Carnahan-Starling equation of state
\cite{Carnahan_Starling_1969} for the elastic hard sphere fluid.
With the exception of the highest density, the compressibility factor
for homogeneously sheared inelastic spheres is quite similar to that
for elastic hard spheres.
The predictions of Enskog theory and Montanero et al.\ for the
pressure (see Fig.~\ref{fig:pressure}a) agree fairly well with the
simulation data.  The main source of the discrepancy is due to the
mis-prediction of the kinetic contribution to the pressure.

The shear viscosity of a granular material is perhaps the most
important design parameter in fast flows, quantifying the power lost
per unit volume. The shear viscosity of the inelastic hard-sphere
system was computed by two means.  The first method is via the
definition of the shear viscosity $\eta$ for simple Couette flow
\begin{equation}
\eta \equiv -\frac{\langle P_{xy}\rangle}{ \dot{\gamma}}
\end{equation}
An alternative method is to perform an energy balance.  The work of
shearing inputs energy into the system.  Collisions between the
inelastic spheres continuously dissipate kinetic energy.  At
steady-state, the average rate of energy input is equal to the average
rate of energy dissipation \cite{Turner_Woodcock_1990}:
\begin{equation}
\eta {\dot{\gamma}}^{2}V 
= -\langle\dot{E}\rangle 
\label{eq:power_loss}
\end{equation}
where $\langle\dot{E}\rangle$ is the average rate of kinetic energy
dissipation.  The rate of energy dissipation is directly related to
the mean time between collisions $t_{\rm avg}$ for a sphere by
\begin{equation*}
\langle\dot{E}\rangle  = \frac{N}{2t_{\rm avg}}
\langle\Delta{E}\rangle
\end{equation*}
where $N$ is the total number of spheres in the system, and
$\langle\Delta{E}\rangle$ is the average amount of kinetic energy
lost per collision.

\begin{figure}
\begin{center}
\includegraphics[clip,width=\columnwidth]{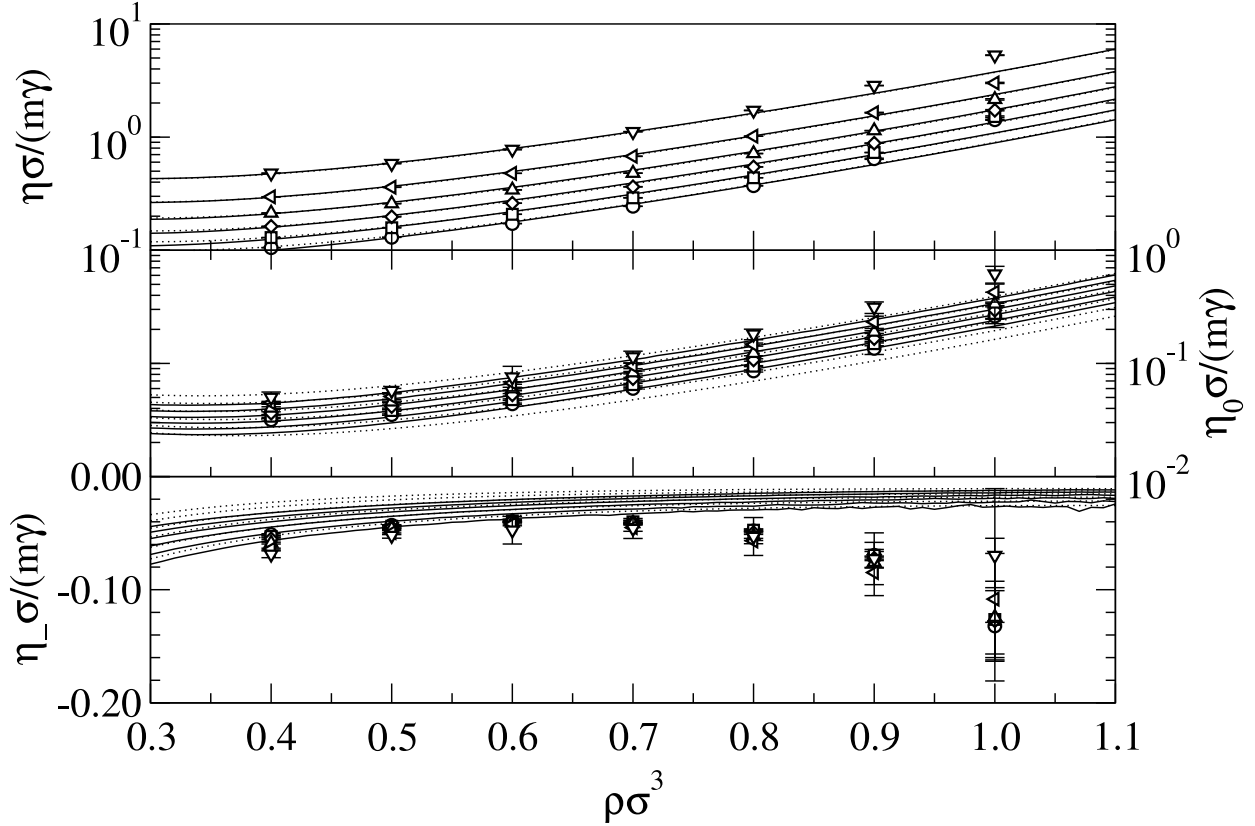}
\caption{\label{fig:eta}
The viscosity for a homogeneously sheared inelastic-hard-sphere system with
(i)   $\alpha=0.4$ (circles),
(ii)  $\alpha=0.5$ (squares),
(iii) $\alpha=0.6$ (diamonds),
(iv)  $\alpha=0.7$ (triangles-up),
(v)   $\alpha=0.8$ (triangles-left), and
(vi)  $\alpha=0.9$ (triangles-down).
The dotted lines are the suggested expressions of
Ref.~\cite{Montanero_etal_1999}, and the solid lines are DSMC
results.}
\end{center}
\end{figure}

The simulation results for the viscosity of sheared inelastic
hard-sphere systems are summarized in Table~\ref{tab:visc}.  The upper
entries are the values obtained from the stress tensor (see
Eq.~(\ref{eq:stress_tensor})), and the lower entries are the values
obtained from the dissipation of kinetic energy (see
Eq.~(\ref{eq:power_loss})).  For all the simulation runs, the two
agree within the statistical uncertainties of the simulations.
Figure~\ref{fig:eta}a shows the dependence of the shear viscosity on
the reduced density of the system, for various values of the
coefficient of restitution.  The viscosity increases with packing
fraction and coefficient of restitution (remembering that the shear
rate is equal to one).  The theory of Montanero et al. captures the
full Enskog behaviour and predicts the viscosity well.  Enskog theory
deviates at low values of $\alpha$ and high densities where the
predictions for the temperature begin to deviate from the simulation
results (see Fig.~\ref{fig:ke}).

In addition to the shear viscosity, we also monitor the in-plane
normal stress coefficient $\eta_-$ and the out-of-plane normal stress
coefficient $\eta_0$, which are defined as \cite{EVANS_MORRIS_1990}
\begin{align*}
\eta_- &= -\frac{1}{2\dot{\gamma}} 
(\langle P_{xx}\rangle-\langle P_{yy}\rangle)
\\
\eta_0 &= -\frac{1}{2\dot{\gamma}}
\left[\langle P_{zz}\rangle
-\frac{1}{2}(\langle P_{xx}\rangle+\langle P_{yy}\rangle)
\right]
\end{align*}
The in-plane normal stress coefficient is plotted in
Fig.~\ref{fig:eta}b, and the out-of-plane normal stress coefficient is
plotted in Fig.~\ref{fig:eta}c.  The simulation values deviate
significantly from the predictions of Enskog theory; however, this is
unsurprising as the velocity dispersion predictions deviate
significantly from the simulation results (see Fig.~\ref{fig:ratio}).

\subsection{Collision statistics}

\begin{figure}
\begin{center}
\includegraphics[clip,width=\columnwidth]{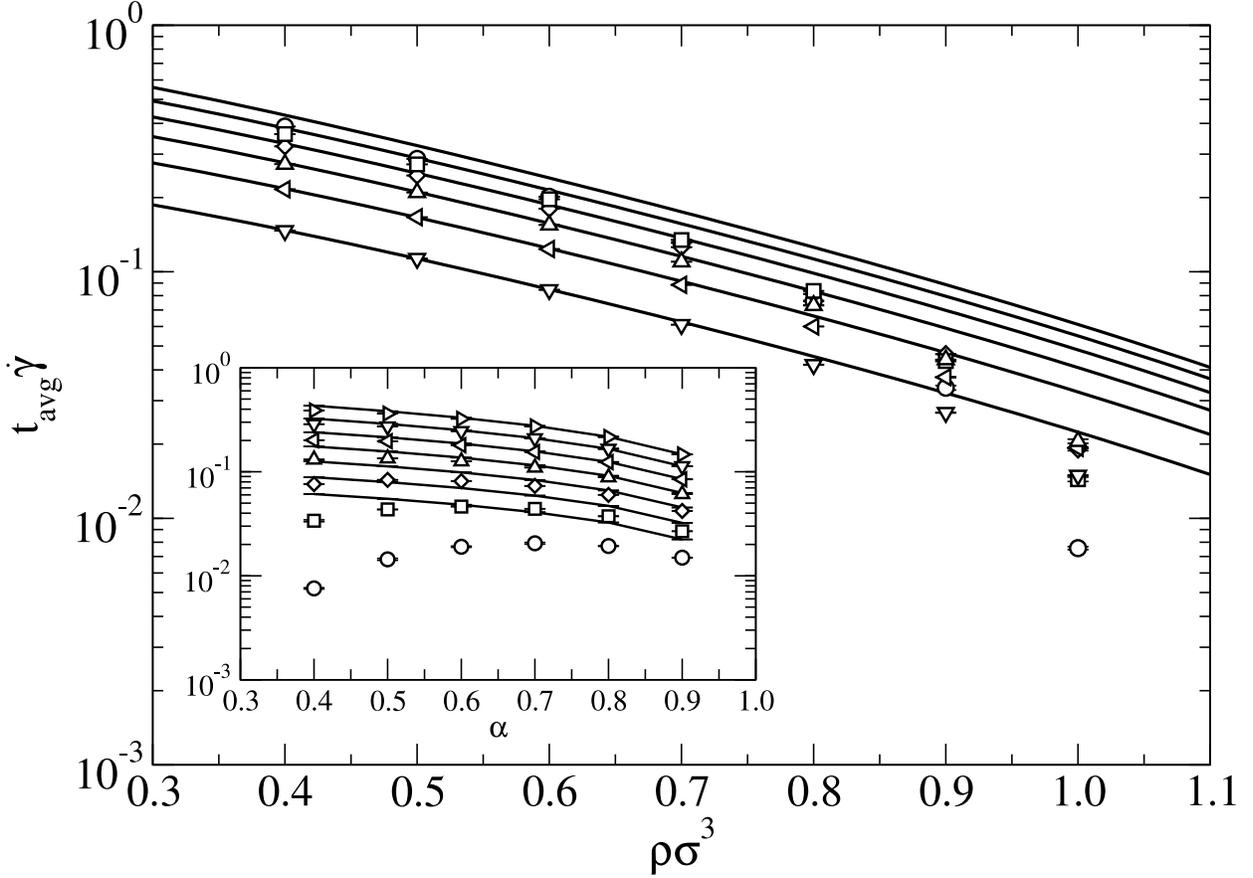}
\caption{\label{fig:t_N864}
Mean time between collisions for sheared inelastic hard spheres with: 
(i)   $\alpha=0.4$ (circles),
(ii)  $\alpha=0.5$ (squares),
(iii) $\alpha=0.6$ (diamonds),
(iv)  $\alpha=0.7$ (triangles-up),
(v)   $\alpha=0.8$ (triangles-left), and
(vi)  $\alpha=0.9$ (triangles-down).
Inset: variation of the mean time between collisions with the coefficient
of restitution for 
(i)   $\rho\sigma^3=1.0$ (circles),
(ii)  $\rho\sigma^3=0.9$ (squares),
(iii) $\rho\sigma^3=0.8$ (diamonds),
(iv)  $\rho\sigma^3=0.7$ (triangles-up),
(v)   $\rho\sigma^3=0.6$ (triangles-left),
(vi)  $\rho\sigma^3=0.5$ (triangles-down), and
(vii) $\rho\sigma^3=0.4$ (triangles-right).
The solid lines are the DSMC simulation results.}
\end{center}
\end{figure}

In this section, we examine the statistics of the collisions
experienced by the spheres.  The mean time between collision
$t_{\rm{}avg}$ provides a characteristic time scale for the sheared
inelastic hard sphere system.  Figure~\ref{fig:t_N864} shows the
variation of the mean time between collisions with the density of the
system at various values of the coefficient of restitution.  The time
between collision decreases with a increasing density, which is
expected; increasing the coefficient of restitution decreases the mean
time between collision.
The variation of $t_{\rm avg}$ with the coefficient of restitution is
given in the inset of Fig.~\ref{fig:t_N864}.  At densities roughly
below $\rho\sigma^3=0.6$, the mean time between collision decreases
monotonically with increasing values of the coefficient of resitution.
However, at higher densities, there is a maximum in $t_{\rm avg}$. The
Enskog theory results describe the results qualitatively well for low
density systems but fail at high densities.

For an elastic fluid, the velocities of different particles are,
in general, uncorrelated.  Consequently, the velocity statistics of
the individual collisions can be determined exactly.  On the other
hand, the particle velocities in a driven granular system can be
strongly correlated, and their on-collision statistics are not exactly
known.

\begin{figure}
\begin{center}
\includegraphics[clip,width=\columnwidth]{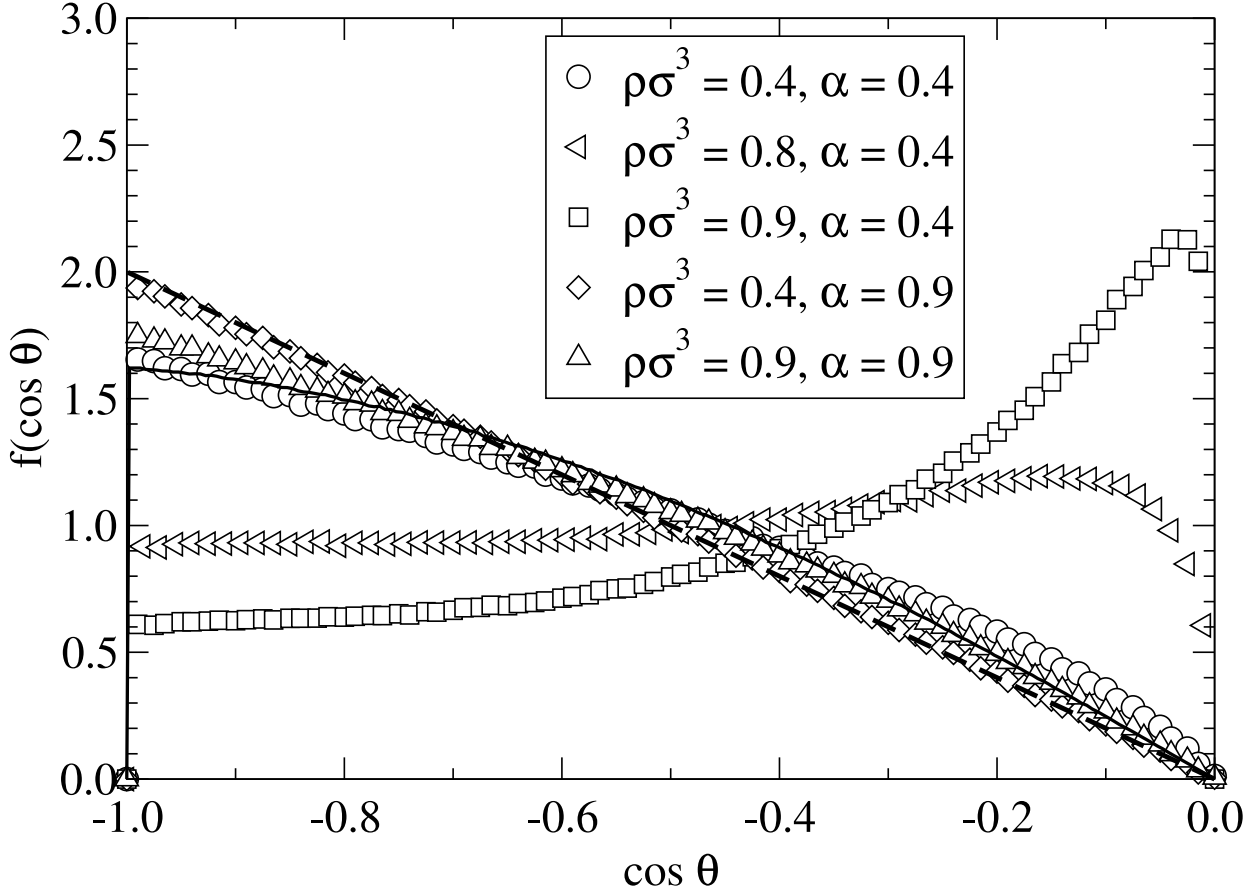}
\caption{\label{fig:cos}
The distribution of collisional angles for 
(i) $\rho\sigma^3=0.4$ and $\alpha=0.4$ (circles), 
(ii) $\rho\sigma^3=0.8$ and $\alpha=0.4$ (triangles-left),
(ii) $\rho\sigma^3=0.9$ and $\alpha=0.4$ (squares),
(iii) $\rho\sigma^3=0.4$ and $\alpha=0.9$ (diamonds), and
(iv) $\rho\sigma^3=0.9$ and $\alpha=0.9$ (triangles-up).
The dashed line is for elastic hard spheres and the solid line is
from a DSMC simulation of $\rho\sigma^3=0.9$ and $\alpha=0.4$.}
\end{center}
\end{figure}

The distribution of the angle $\theta$ between the relative velocity
and the relative position of two spheres on collision
($\cos\theta={\bf{}r}_{ij}\cdot{\bf{}v}_{ij}\,/\left|{\bf{}r}_{ij}\right|\left|{\bf{}v}_{ij}\right|$)
is given in Fig.~\ref{fig:cos}.  The solid line denotes an isotropic
collision distribution (as is the case for elastic,
elastic-hard-sphere systems).  The symbols are the simulation data for
sheared inelastic hard spheres, and the solid line is the DSMC result
for $\rho\sigma^3=0.9$ and $\alpha=0.4$.  For weakly inelastic
systems, the distribution of the collisional angle is close to that
for the elastic hard-sphere system.  As the inelasticity and
density of the particles increases, however, there is a gradual
increase of the frequency of ``glancing'' collisions (where
$\cos\theta$ is near $0$) at the expense of more ``head-on''
collisions (where $\cos\theta$ is close to $-1$).  This is in
agreement with the two-dimensional shearing simulation of Tan and
Goldhirsch \cite{Tan_Goldhirsch_1997} and Campbell and Brennen
\cite{Campbell_Brennen_1985}.  The Enskog theory does not capture this
effect, as the DSMC simulations only display a small increased bias
towards glancing collisions even in the dense, highly inelastic
system.

The increase in glancing collisions for strongly inelastic systems
(see Fig.~\ref{fig:cos}) results primarily from collisions between
pairs of particles orientated in the $x$-$y$ plane.  This occurs when
the change of the streaming velocity over the diameter of a particle
becomes significant in comparison to the average relative peculiar
velocity \cite{Alam_Luding_2003}.  Particles separated in the
$y$-plane then have a significantly increased relative velocity which
increases their probability of collision. Both the DSMC and granular
dynamics simulation results support this; however, DSMC does not
exhibit the large increase in collisions with a very large collision
angle.

\begin{figure}
\begin{center}
\includegraphics[clip,width=\columnwidth]{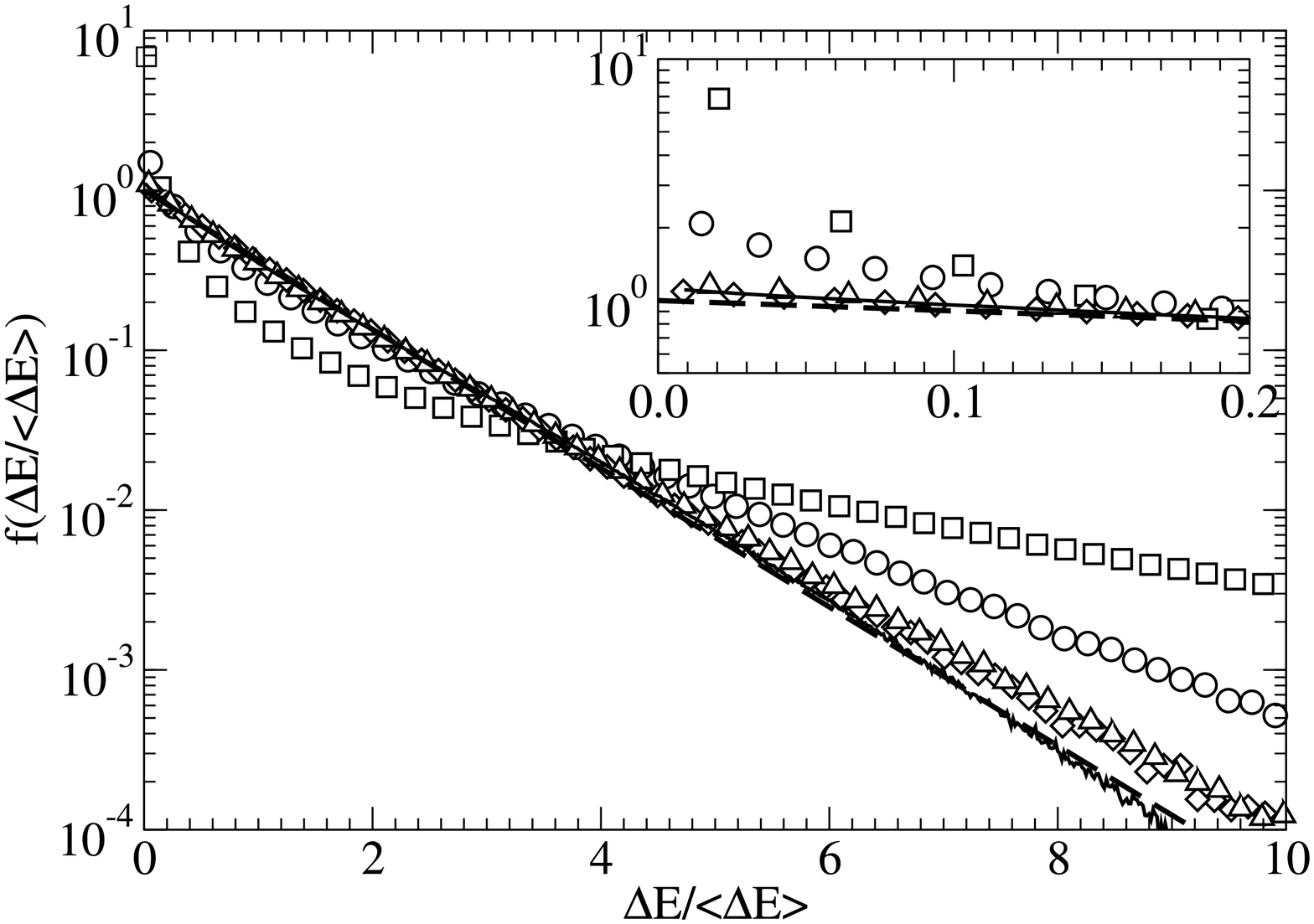}
\caption{\label{fig:ke_coll}
The change of kinetic energy on collision for sheared inelastic hard
sphere systems with
(i) $\rho\sigma^3=0.4$ and $\alpha=0.4$ (circles), 
(ii) $\rho\sigma^3=0.9$ and $\alpha=0.4$ (squares),
(iii) $\rho\sigma^3=0.4$ and $\alpha=0.9$ (diamonds), and
(iv) $\rho\sigma^3=0.9$ and $\alpha=0.9$ (triangles).
The dashed line represents the kinetic energy loss on collision if the
velocity were given by the Maxwell-Boltzmann distribution, and the
solid line is from a DSMC simulation at $\rho\sigma^3=0.9$ and
$\alpha=0.4$.  The inset highlights the frequency of collisions that
result in low energy losses.}
\end{center}
\end{figure}

In the inelastic hard sphere system, every collision results in a loss
of kinetic energy.  The simulation results for the distribution of the
loss of kinetic energy on collision is given in
Fig.~\ref{fig:ke_coll}.  If the velocity distribution of the spheres
were Gaussian (e.g., Maxwell-Boltzmann distribution), then the kinetic
energy loss on collision would be distributed according to a Poisson
distribution:
\begin{equation*}
f(\Delta{E}) = \frac{1}{\langle\Delta{E}\rangle}
\exp\left(-\frac{\Delta{E}}{\langle\Delta{E}\rangle}\right)
\end{equation*}
This is given by the solid line in Fig.~\ref{fig:ke_coll}.  At high
values of $\alpha$, the distribution of the change of kinetic energy
on collision is nearly exponential; for these systems, density does
not significantly affect the results.  

As $\alpha$ decreases, there is a greater frequency of collisions that
result in a very slight loss of kinetic energy (i.e.\ the initial peak
in Fig.~\ref{fig:ke_coll}).  This corresponds to the increase in the
glancing collisions in the systems.
This enhancement of relatively elastic collisions is accompanied by an
increase in collisions that result in large losses of kinetic energy
(i.e.\ the long tail in Fig.~\ref{fig:ke_coll}).  These result from
``head-on'' collisions, which occur between particles oriented
primarily in the $x$-direction where the velocity dispersion is the
greatest.  While these ``head-on'' collisions occur less frequently
than glancing collisions in the highly inelastic systems, they are
more violent.
Increasing the density enhances these effects.

Thus far, we have only studied the statistics of single collisions.
One common assumption in many kinetic theories is that the individual
collisions experienced by a particle are statistically independent.
We now study the correlation between collisions by examining the time
required for a particle to undergo a number of collisions.  If the
various collisions experienced by a particle can be considered to
arrive at random times in an independent manner, then the time $t$
required for a particle to undergo $n$ collisions is given by a
Poisson process.  The probability density function $p_n(t)$ that a
particle experiences $n$ collisions in a period of time $t$ is
\begin{equation}
p_n(t) = \frac{(t/t_{\rm avg})^{n-1}}{t_{\rm avg}\Gamma(n)}
\exp\left(-\frac{t}{t_{\rm avg}}\right)
\label{eq:Poisson}
\end{equation}
where $\Gamma(n)$ is the Gamma function.  Deviations from this
distribution are an indication of correlations between collisions.
For elastic hard-sphere fluids, the Poisson process describes the
collision time distribution fairly well, however, there are noticeable
deviations, even at low densities, which increase with increasing
density \cite{Lue_2005,Visco_etal_2008a,Visco_etal_2008b}..

\begin{figure}
\begin{center}
\includegraphics[clip,width=\columnwidth]{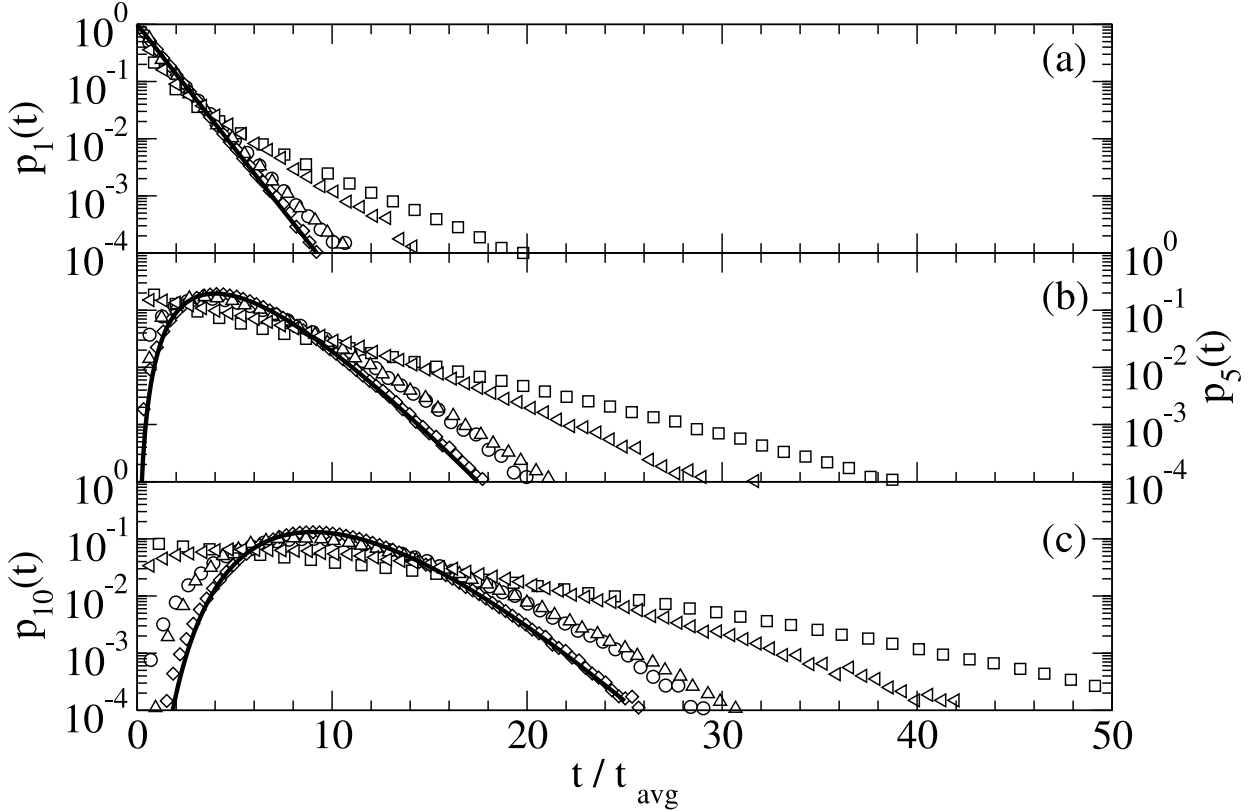}
\caption{\label{fig:foft_1coll}
Distribution of time between 
(a) one collision,
(b) five collisions, and
(c) ten collisions
in sheared inelastic hard spheres with
(i) $\rho\sigma^3=0.4$ and $\alpha=0.4$ (circles), 
(ii) $\rho\sigma^3=0.9$ and $\alpha=0.4$ (squares),
(iii) $\rho\sigma^3=0.4$ and $\alpha=0.9$ (diamonds), and
(iv) $\rho\sigma^3=0.9$ and $\alpha=0.9$ (triangles).
The solid line is for a Poisson process.
}
\end{center}
\end{figure}

The collision time distributions for homogeneously sheared inelastic
hard sphere systems are shown in Fig.~\ref{fig:foft_1coll}.  The solid
lines denote the Poisson distribution, given by
Eq.~(\ref{eq:Poisson}).  
At high values of the coefficient of restitution, the distributions
are similar to those of elastic hard spheres and are fairly well
described by a Poisson process.  As the coefficient of restitution
decreases, however, the simulation data deviate significantly from the
Poisson process, indicating very strong correlations between
collisions.  Qualitatively, the deviations are similar to that
observed for elastic hard sphere systems: there is an enhancement of
very short and very long wait-times between collisions.  However,
these differences are much more pronounced for the inelastic hard
sphere systems.

\section{Conclusions
\label{sec:conclusions}}

We have performed non-equilibrium molecular dynamics simulations of
sheared inelastic-hard-systems using the SLLOD algorithm combined with
Lees-Edwards boundary conditions.  In these simulations, care was
taken to ensure that the systems remain homogeneous and the shear was
uniform across the system.  As a consequence, these simulations may
prove a useful reference to compare with the predictions of kinetic
theory.

DSMC simulations of the Enskog equation were performed to provide a
solution to the kinetic theory without further approximation.  These
compared favorably with the simulation results except for dense,
strongly inelastic systems.  The velocity anisotropy effect can be
very strong even in homogeneous systems, and kinetic theory solutions
must take this into account in their approximations.

Results were presented for the velocity statistics of individual
particles in the system.  The velocity distributions were, in general,
well described by an anisotropic Gaussian.  Theories based on the
anisotropic Gaussian and the full second moment balance (e.g., see
Ref.~\cite{Chou_Richman_1998}) are well suited to these systems.
Sheared, inelastic-hard-sphere systems do not obey the equipartition
theorem.  Fluctuations of the velocity in the $x$-direction (the
direction of shear) were greater than those in the $y$- and
$z$-directions, which were both similar to each other.  In addition,
the granular temperature, which characterizes the overall fluctuation
of the velocity, was observed to possess a minimum with respect to the
density.  This minimum becomes more pronounced as the coefficient of
restitution of the spheres decreases.

The variation of the stress in the system was also examined.  The
compressibility factor of the sheared inelastic-hard-sphere system was
quite similar to that of elastic hard spheres, as estimated by the
Carnahan-Starling equation of state.  The shear viscosity of the
systems was computed in two different manners: from the average of the
stress tensor and from the rate of dissipation of kinetic energy.  The
value of the viscosity from both these methods agree to within the
statistical uncertainty of the simulations.  The predictions of the
Enskog equation and the kinetic theory of Montanero et al.\
\cite{Montanero_etal_1999} were in fairly good agreement with the
simulation data.  The in-plane and out-of-plane stress coefficients
were also computed, but the kinetic theory predictions for these
quantities were not as accurate.

Finally, the collision statistics of particles in the sheared
inelastic hard sphere system were studied.  The mean time between
collision was found to decrease monotonically with increasing density;
however, at fixed density, it displays a maximum at intermediate
values of the coefficient of restitution.  Examination of the
collision time distributions indicated the presence of strong
correlations between collisions.  Including these correlations within
a kinetic theory will be important in developing an accurate
description of high density, sheared inelastic-hard-sphere systems.

\bibliography{main,inelastic2}
\bibliographystyle{apsrev}

\begin{table*}[htb]
\begin{center}
  \caption{\label{tab:visc} Dimensionless viscosity
    $\eta\sigma/(m\dot{\gamma})$ of sheared inelastic hard-sphere
    systems at various densities $\rho$ and coefficients of
    restitution $\alpha$.  The upper value is determined from the stress
    tensor (see Eq.~(\ref{eq:stress_tensor})), and the lower value is
    determined from the kinetic energy dissipation rate (see
    Eq.~(\ref{eq:power_loss})). The value in brackets is the standard
    deviation over all of the runs.}
\begin{tabular}{|c|c|c|c|c|c|c|}
\hline
\backslashbox{$\rho\sigma^3$}{$\alpha$} & 0.4 & 0.5 & 0.6 & 0.7 & 0.8 & 0.9 \\
\hline\hline
0.4 & 0.1054(0.0003) & 0.1294(0.0004) & 0.1625(0.0006) & 0.2134(0.0005) & 0.2970(0.0004) & 0.480(0.002) \\
 & 0.1054(0.0003)& 0.1294(0.0004)& 0.1625(0.0006)& 0.2134(0.0005)& 0.2969(0.0004)& 0.480(0.002)\\
\hline
0.5 & 0.1300(0.0003) & 0.1582(0.0002) & 0.1979(0.0007) & 0.2585(0.0003) & 0.361(0.001) & 0.583(0.002) \\
 & 0.1300(0.0003)& 0.1582(0.0002)& 0.1979(0.0007)& 0.2585(0.0003)& 0.361(0.001)& 0.583(0.001)\\
\hline
0.6 & 0.1722(0.0005) & 0.2078(0.0003) & 0.2606(0.0007) & 0.3416(0.0009) & 0.4785(0.0005) & 0.779(0.006) \\
 & 0.1722(0.0005)& 0.2078(0.0003)& 0.2606(0.0007)& 0.3417(0.0009)& 0.4785(0.0007)& 0.779(0.007)\\
\hline
0.7 & 0.2445(0.0004) & 0.2909(0.0008) & 0.364(0.001) & 0.4801(0.0008) & 0.677(0.002) & 1.117(0.005) \\
 & 0.2446(0.0003)& 0.2909(0.0008)& 0.364(0.001)& 0.4801(0.0009)& 0.677(0.002)& 1.117(0.005)\\
\hline
0.8 & 0.371(0.001) & 0.4375(0.0002) & 0.545(0.001) & 0.716(0.003) & 1.020(0.002) & 1.722(0.008) \\
 & 0.371(0.001)& 0.4375(0.0002)& 0.545(0.001)& 0.716(0.003)& 1.020(0.002)& 1.722(0.008)\\
\hline
0.9 & 0.643(0.007) & 0.731(0.001) & 0.885(0.002) & 1.141(0.004) & 1.64(0.01) & 2.86(0.01) \\
 & 0.643(0.008)& 0.731(0.002)& 0.885(0.002)& 1.141(0.004)& 1.64(0.01)& 2.856(0.010)\\
\hline
1.0 & 1.42(0.01) & 1.52(0.02) & 1.73(0.02) & 2.15(0.03) & 3.01(0.03) & 5.30(0.04) \\
 & 1.42(0.01)& 1.52(0.02)& 1.73(0.02)& 2.15(0.03)& 3.01(0.04)& 5.30(0.03)\\
\hline
\hline
\end{tabular}
\end{center}
\end{table*}
\end{document}